\documentclass[aps,pra,twocolumn, amsmath, amssymb,superscriptaddress]{revtex4}
\usepackage{graphicx}
\usepackage{dcolumn}
\usepackage{bm}
\usepackage{indentfirst}

\begin{document}

\title{Ground state EIT cooling of $^{171}$Yb$^+$ ion}

\author{D. S. Krysenko}
\affiliation{Institute of Laser Physics, 630090, Novosibirsk, Russia}
\affiliation{Novosibirsk State Technical University, 630073, Novosibirsk,
Russia}

\author{O.N. Prudnikov}
\email{oleg.nsu@gmail.com}
\affiliation{Institute of Laser Physics, 630090, Novosibirsk, Russia}
\affiliation{Novosibirsk State University, 630090, Novosibirsk, Russia}

\author{A. V. Taichenachev}
\affiliation{Institute of Laser Physics, 630090, Novosibirsk, Russia}
\affiliation{Novosibirsk State University, 630090, Novosibirsk, Russia}

\author{V. I. Yudin}
\affiliation{Institute of Laser Physics, 630090, Novosibirsk, Russia}
\affiliation{Novosibirsk State University, 630090, Novosibirsk, Russia}
\affiliation{Novosibirsk State Technical University, 630073, Novosibirsk,
Russia}

\author{S. V. Chepurov}
\affiliation{Institute of Laser Physics, 630090, Novosibirsk, Russia}

\author{S. N. Bagaev}
\affiliation{Institute of Laser Physics, 630090, Novosibirsk, Russia}

\date{\today}

\begin{abstract}
We propose a scheme of deep laser cooling of $^{171}$Yb$^{+}$, which is based on the effect of electromagnetically induced transparency (EIT) in a polychromatic field with three frequency components  resonant with optical transitions of the $^2S_{1/2} \to \, ^2P_{1/2}$ line. The deep cooling down to the ground motional state in a trap allows for a significant suppression of the second order Doppler shift in frequency standard. Moreover, in our scheme, there is no need to use a magnetic field, which is required for Doppler cooling of $^{171}$Yb$^{+}$ in a field with two-frequency component. The cooling without use of magnetic field is important for deep suppression of quadratic Zeeman shifts of clock transitions due to uncontrolled residual magnetic field.
\end{abstract}


\maketitle


\section{Introduction}
Laser cooling is a necessary step for modern experiments with quantum systems based on neutral atoms and ions that have wide applications, including quantum metrology, the study of fundamental properties of cold atomic Bose and Fermi condensates \cite{Cornell2002,Ketterle,DeMarco,DeMarco_PRL}, implementation of quantum logic elements and quantum computing \cite{Nielsen}. The development of modern frequency standards using cold atoms \cite{Falke,Katori2020,McGrew} and ions \cite{Wineland_Al,Huntemann,Huang_Ca} 
has become highly relevant. The achieved level of accuracy and long-term stability of optical frequency standards at the level $10^{-18}$, opens up new horizons for modern fundamental researches, such as the study of the effects of Earth's gravitation on the space-time continuum \cite{Katori2020,Lion2017,Ludlow2018}, the test of fundamental constants \cite{Godun14,Huntemann14}, verification of the general relativity, Lorentz invariance of space \cite{Dzuba,Sanner,Laura}, the search for dark matter \cite{Arvanitaki,Stadnik}, etc.

\begin{figure}[b]
\centerline{\includegraphics[width=1.1 in]{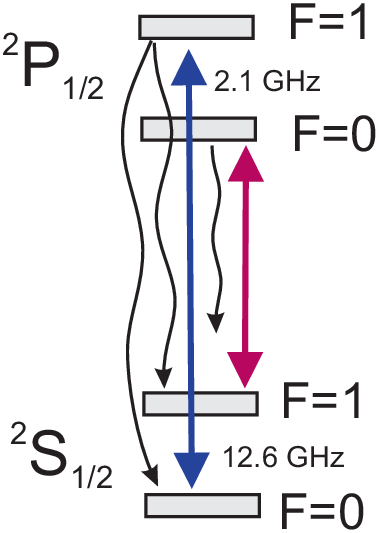}}
\caption{The energy level of the hyperfine structure $^2S_{1/2}$ and $^2P_{1/2}$ in $^{171}$Yb$^{+}$ ion, which are used for laser cooling. The solid lines represent the transitions induced by two-frequency components of the field. The wavy lines represent the main channels of spontaneous decay. The magnetic field is required here to destroy the CPT effect (coherent trap state) on the $^2S_{1/2}(F=1)$ level via $^2S_{1/2}(F=1) \to \,^2P_{1/2}(F=0)$ optical transition.} \label{fig:F1}
\end{figure}

To achieve high precision level of frequency standards, it is necessary to take into account systematic frequency shifts of different nature. Therefore, works are aimed at suppression of these shifts are very important. For example, in the context of the $^{171}$Yb$^+$ ion-based frequency standard, further progress can be linked to the control and suppression of systematic shifts caused by residual magnetic field, black-body radiation (BBR) shifts, and  quadratic Doppler shifts \cite{Huntemann, Sanner}. However, the main challenge here is that the transition $^2S_{1/2}\,$$\to$$\,^2P_{1/2}$ is used for laser cooling is not closed, that require to use a laser field  with at least two frequency components \cite{Tamm, Prudnikov_2017, Prudnikov_2019} (see Fig.\ref{fig:F1}). In this case, a  relatively large magnetic field  of $\sim$1-10 G is required to destroy the coherent trap state at the $^2S_{1/2}(F=1)$ level via $^2S_{1/2}(F=1) \to \,^2P_{1/2}(F=0)$ optical transition due to coherent population trapping (CPT) effect. The laser cooling here can reach minimum temperature that corresponds to the Doppler limit $k_B\, T_D \simeq \hbar \gamma/2$, where $\gamma$ is the natural linewidth of the optical transition $^2S_{1/2} \to \, ^2P_{1/2}$. 
The hysteresis effects during the switching off the magnetic field is required for cooling create certain difficulties in minimizing the residual magnetic field and keeping it constant in various cooling and clock operation cycles. The similar difficulties arise when implementing quantum logic and quantum computing elements based on $^{171}$Yb$^+$ ions \cite{Kolachevsky}.

In this paper, we propose an alternative method of laser cooling that makes it possible to eliminate the use of a magnetic field and, in contrast to the standard scheme \cite{Tamm,Prudnikov_2017,Prudnikov_2019}, allows atoms to be cooled significantly below the Doppler limit $T_D$ and, thus, significantly suppress the second-order Doppler shift in a frequency standard.

\section{Deep laser cooling of $^{171}\mbox{Yb}^{+}$}
For laser cooling of $^{171}$Yb$^{+}$ ion, the light field with at least two frequency components have to be used \cite{Tamm,Prudnikov_2017,Prudnikov_2019} (see Fig.\,\ref{fig:F1}). Here, one of the frequency components is close to the $^2S_{1/2} (F=0) \to \, ^2P_{1/2}(F=1)$ transition, and the other one to the $^2S_{1 /2} (F=1) \to \, ^2P_{1/2}(F=0)$ transition. 
The laser cooling arises as a result of the action of the dissipative Doppler force on a moving ion, which leads to cooling only to the temperature of the Doppler limit $k_B T_D = \hbar \gamma/2$. In this scheme, an additional magnetic field is required to  
destroy the coherent trap state at the $^2S_{1/2}(F=1)$ level via $^2S_{1/2}(F=1) \to \,^2P_{1/2}(F=0)$ optical transition due to CPT effect \cite{Prudnikov_2017,Prudnikov_2019}.

Deeper laser cooling of ions, down to the ground motional  state, can be achieved under conditions of resolved sideband cooling \cite{Wineland,Javanainen1981,Wineland2003}, when the ion is localized in a trap on scales smaller than the wavelength (the Lamb-Dicke parameter $\eta = \sqrt{E_R/\hbar\, \omega_{osc}} \ll 1$, where $E_R = \hbar^2 k^2/2M$ is the recoil energy, $M$ is the ion mass), and $\omega_{osc}$ is the ion oscillations frequency in the trap is enough large, $\omega_{osc} \gg \gamma$ (i.e., transitions between different motional states of the ion have to be spectrally resolved). However, these conditions are not satisfied for  $^2S_{1/2} \to \, ^2P_{1/2}$ cooling transition in $^{171}$Yb$^{+}$ ion, where the natural line width $\gamma/2\pi = 23\, $~MHz and the typical oscillation frequency in the trap  $\omega_{osc}/2\pi \simeq 400 - 600\, $~kHz.

The laser cooling by  using the electromagnetically induced transparency (EIT) technique \cite{Morigi2000}  do not require condition $\omega_{osc} \gg \gamma$. To implement it, a three-level $\Lambda$ system is required, in which transitions are induced by a pair of light waves. Under conditions when the detunings of light waves are equal, the atoms are pumped into a dark state, which  does not interact with the field. This  allows to substantially suppress the heating processes associated with the emission of spontaneous photons.
Moreover, the presence of a narrow EIT resonance, with a width much smaller than spontaneous decay rate $\gamma$ of excited state, enables cooling via two-photon transitions between different motional states of the ground levels in the $\Lambda$-scheme, similar to the Raman cooling technique  \cite{Morigi2000, Morigi2003, Roos2016}.

The choice of the interaction scheme for the implementation of the EIT cooling of $^{171}$Yb$^{+}$ ion is a non-trivial task. In Ref.\,\cite{Khabarova}, it was proposed to use three frequency components near the resonance of the optical transition $^2S_{1/2} (F=1) \to \,^2P_{1/2}(F=0)$, known as the double EIT scheme \cite{Evers}. In this case, each frequency component determines transitions between different Zeeman levels of the ground state $|^2S_{1/2}, F=1, \mu = 0, \pm 1 \rangle$ and the excited state $|^2P_{1/2}, F=0, \mu = 0 \rangle$. In addition, to repump from $^2S_{1/2} (F=0)$ state it is necessary the use of an extra laser field resonant to  $^2S_{1/2} (F=0) \to \,^2P_{1/2}(F=1)$ transition. Thus the laser cooling scheme \cite{Khabarova} requires four frequency components, that significantly complicates the overall scheme of laser cooling. As well, in the recent paper  \cite{Qiao} similar double EIT scheme for laser cooling of $^{171}$Yb$^+$ was theoretically and experimentally investigated. In compare to \cite{Khabarova}, there only two frequency components resonant to $^2S_{1/2}(F=1) \to \,^2P_{1/2}(F=0)$ transition are used, but it also requires a magnetic field to destroy the coherent trap state at the $^2S_{1/2}(F=1)$ level. For both these cases additional optical pumping have to be carried out to prepare the initial clock state $^2S_{1/2}(F=0)$.

To implement deep EIT cooling, as well as the preliminary Doppler cooling preceding it, we propose to use a polychromatic field of running waves with only three frequencies
\begin{equation}\label{field}
    {\bf E}({\bf r},t) = Re\left\{ \sum_{p =1,2,3} {\bf E}_p e^{i {\bf k_p} {\bf r}} e^{-i\omega_p t} \right\},
\end{equation}
where ${\bf E}_p$ are complex vectors, which define the polarization and amplitude of each frequency component  $p = 1,2,3$.

\begin{figure}[t]
\centerline{\includegraphics[width=3.2 in]{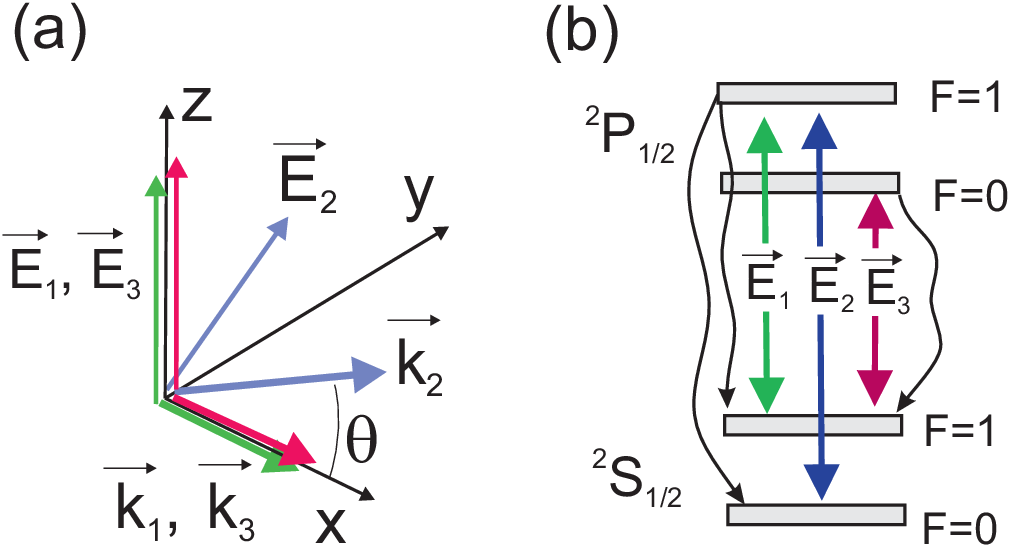}}
\caption{The three-frequency field configuration formed by three running waves (a).
The transitions induced by each frequency components (b).} \label{fig:F2}
\end{figure}

In our configuration, the field components have linear polarizations [see Fig.\,\ref{fig:F2}(a)]. The wave vectors of the ${\omega}_1$ and ${\omega}_3$ frequency components are directed along the $x$-axis, and their polarization vectors ${\bf E}_1$ and ${\bf E}_3$ are along the $z$-axis. The wave vector of the ${\omega}_2$ component lies in the $(xz)$-plane with some angle $\theta$ to  $x$-axis. The orientation of the polarization vector ${\bf E}_2$ is varied for different cooling stages. The frequencies $\omega_n$ are chosen so that they provide light-induced transitions between different hyperfine levels of the $^2S_{1/2}$ and $^2P_{1/2}$ states according to the scheme in Fig.\,\ref{fig:F2}(b).

\subsection{Doppler cooling}
\begin{figure}[t]
\centerline{\includegraphics[width=3.2 in]{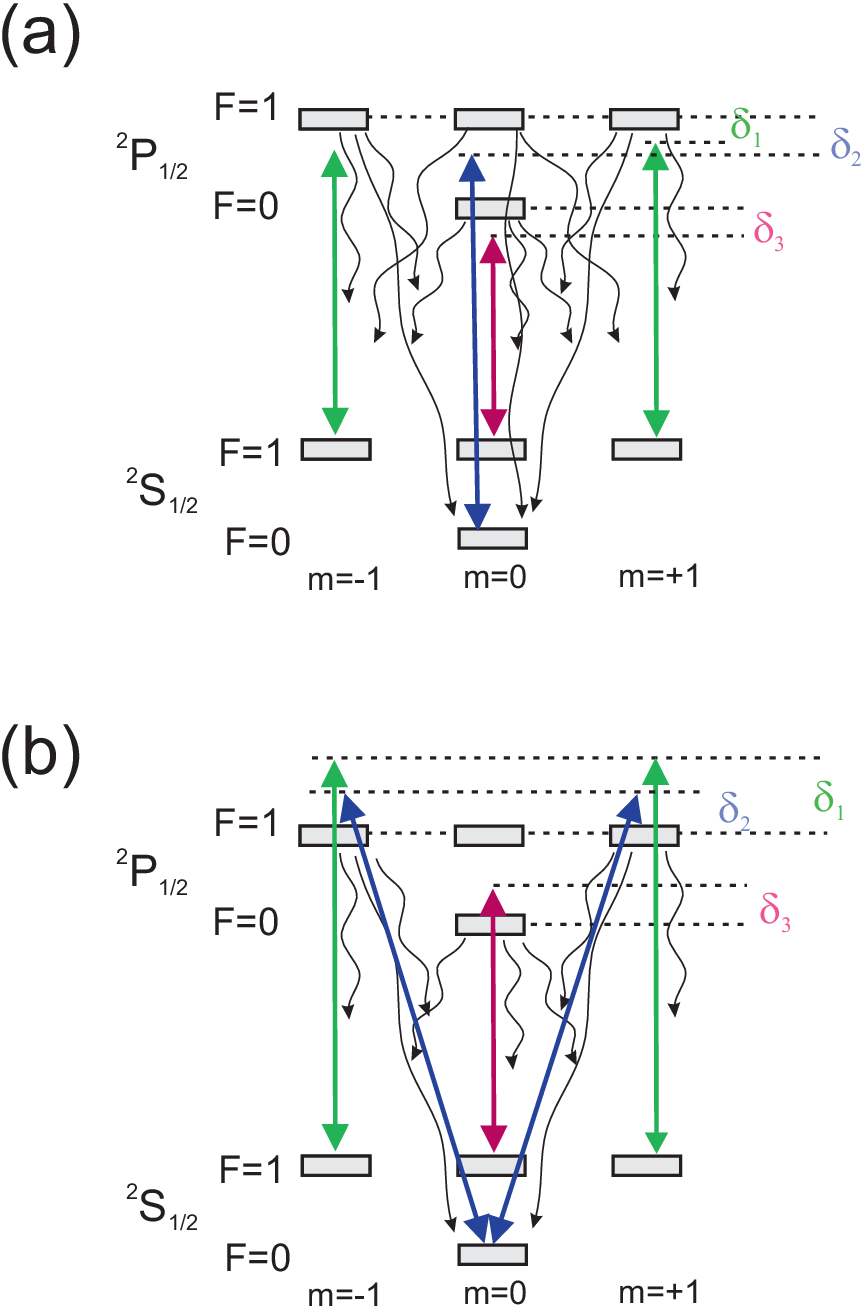}}
\caption{The Zeeman sublevels of hyperfine states $^2S_{1/2}$ and $^2P_{1/2}$ of $^{171}$Yb$^{+}$ and transitions induced by frequency component of the light field for two cooling stages: first stage of Doppler cooling (a) and the second stage of EIT cooling (b). The transitions induced by the frequency components are indicated by double arrows: green -- transitions caused by ${\bf E}_1$ component, blue -- transitions caused by ${\bf E}_2$ component and red -- transitions caused by ${\bf E}_3$ component of the light field. The wavy arrows indicate transitions caused by spontaneous decays.
Here $\delta_1$, $\delta_2$, $\delta_3$ are the corresponding detunings for the frequency components.} \label{fig:F3}
\end{figure}

As the EIT ground-state cooling technique is applicable to ions already prepared at low temperatures \cite{Morigi2000,Morigi2003,Roos2016}, the preliminary Doppler cooling is required. For the considered field configuration (\ref{field}), the effective Doppler cooling can be realized with linear codirectional polarizations of the field components, ${\bf E}_1 \, ||\, {\bf E}_2\,||\, {\bf E}_3$. In this case, the corresponding scheme of resonant light-induced transitions over the Zeeman sublevels is shown in Fig.\,\ref{fig:F3}(a). In Ref.\,\cite{Krysenko2023}, we carried out a detailed analysis of laser cooling in such a field. It was shown that the minimum temperature corresponds to the Doppler  limit
\begin{equation}
k_B T = \hbar \gamma/3 \, ,
\end{equation}
and is achieved for the low intensities under the condition that the Rabi frequencies of each frequency component are equal $\Omega_1 = \Omega_2 = \Omega_3$ ($\Omega_p = |{\bf E}_p|d/\hbar$, $d$ is the transition dipole moment $^2S_{1/2} \to \,^ 2P_{1/2}$), and the detunings of each frequency component have to be chosen
\begin{equation}
\delta_1=\delta_2=\delta_3 = -\gamma/2 \, .
\end{equation}
Here, the detunings are defined as $\delta_p = \omega_p-\omega_{0p}$ the difference between the frequency of the p-th field component ${\bf E}_p$ and the frequency of the corresponding resonant transition $\omega_{0p}$ (see Fig.\,\ref{fig:F3}).

\subsection{Ground state EIT cooling}
For the second stage of deep laser cooling in proposed field configuration Fig.\,\ref{fig:F2}(a), we direct the polarization vector ${\bf E}_2$ along the $y$ axis so that ${\bf E}_2 \perp { \bf E}_{1,3}$. In this case, the interaction with light components forms a double $\Lambda$ scheme for the Zeeman sublevels of hyperfine states $^2S_{1/2}$ and $^2P_{1/2}$ [see Fig.\,\ref{fig:F3}(b)] that allows to implement EIT ground state cooling technique. For EIT cooling in the considered scheme, it is necessary to choose the detuning of the driven field component ${\bf E}_1$ to be blue-detuned, $\delta_1 > 0$. The intensity of this component have to be chosen so that ac Stark shift  of dressed Zeeman sublevels  $|^2S_{1/2}, F=1 , m=\pm 1 \rangle$ are light-shifted upwards by an amount equal to the trap frequency $\omega_{osc}$. The field components ${\bf E}_2$ and ${\bf E}_1$ drive two-photon transitions between the ground states $|^2S_{1/2}, F=1 , m=\pm 1 \rangle$ and $|^2S_{1/2}, F=1 , m=0 \rangle$. For the condition $\delta_2 = \delta_1$, the efficient EIT cooling down to the ground motional state can be achieved, similar to thee-level $\Lambda$ atomic system \cite{Morigi2000,Morigi2003,Roos2016}. The third frequency component ${\bf E}_3$ plays the role of optical pumping for depopulating the state $|^2S_{1/2}, F=1 , m=0 \rangle$. 

The dynamics of the average vibrational quantum number ${\bar N} = \sum_{n=0}^{\infty} n P_n$ (where $P_n$ is the population of the ion's n-th motional state in the trap) is determined by the rate balance equation \cite{Wineland2003,Roos2016}
\begin{equation} \label{nbar}
    \frac{d}{dt}{\bar N} = -\left(A_- - A_+ \right){\bar N} +A_+ \, ,
\end{equation}
where the rate coefficients $A_{\pm}$ in accordance to Ref.\,\cite{Wineland2003}
are given by scattering rate $W(\Delta)$ as a function of two-photon detuning for atom at rest $\Delta = \delta_2-\delta_1$. The scattering rate can be expressed trough the steady-state population $\rho^{ee}$ in the excited state $^2P_{1/2}$  
\begin{equation}
W(\Delta) = \gamma\, \rho^{ee} \, .
\end{equation}
Thus the rate coefficients can be expressed as
\begin{equation}
    A_{\pm} = \eta^2 \left[ W_0 + W_{\mp}\right] \, ,
\end{equation}
where 
\begin{equation}
W_{0} = W(0) \, ,
\end{equation}
and 
\begin{equation}
W_{\pm} = W(\pm\,\omega_{osc}) \, .
\end{equation}
The ion cooling is achieved under conditions $A_->A_+$. In this case, the stationary solution of the equation (\ref{nbar}) has the form:
\begin{equation}\label{Nff}
  {\bar N}_{f} = \frac{A_+}{A_--A_+}=\frac{W_0+W_-}{W_+-W_-}  \, ,
\end{equation}
and determines the minimum laser cooling temperature of the ion in the trap.
The Lamb-Dicke parameter for two-photon transitions is determined by the difference between the wave vectors ${\bf k}_1$ and ${\bf k}_2$
\begin{eqnarray}\label{LD2}
\eta &= &\left|\left({\bf k}_1-{\bf k}_2 \right) \cdot {\bf e}_m \right| \sqrt{\frac{\hbar}{2M\omega_{osc}^{(m)}}} \, \\
&\approx& \left|\left({\bf n}_1-{\bf n}_2 \right) \cdot {\bf e}_m \right| \sqrt{\frac{\hbar k^2}{2M\omega_{osc}^{(m)}}}\,, \nonumber
\end{eqnarray}
where ${\bf e}_m$ denotes the unit vector describing the oscillation direction of the spatial mode to be cooled and $\omega_{osc}^{(m)}$ is the oscillation frequency of this spatial mode \cite{Morigi2000}, ${\bf n}_j = {\bf k}_j/|{\bf k}_j|$ ($j=1,2$), where we assume $|{\bf k}_1| \simeq |{\bf k}_2|=k$. Thus, by varying the angle $\theta$ between the wave vectors ${\bf k}_1$ and ${\bf k}_2$ (see Fig.\,\ref{fig:F2}), as well as the overall direction of the waves relative to the principal axes of the trap, it is possible to control the laser cooling rate of different motional modes.

\begin{figure}[t]
\centerline{\includegraphics[width=3.2 in]{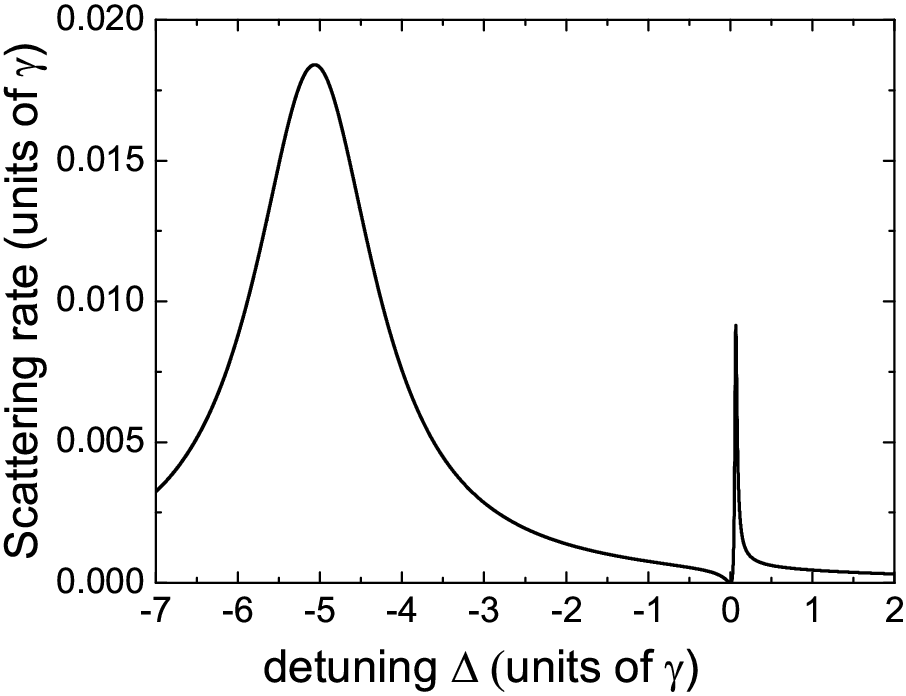}}
\caption{The scattering rate $W(\Delta)$ as function of two-photon detuning. The field parameters are $\delta_1 = 5 \gamma$, $\Omega_1 = 2 \gamma$, $\Omega_2 = 0.2 \gamma$, $S_3 = 0.5$ } \label{fig:F4}
\end{figure}

The expressions for the decay rates $W_0$, $W_+$, and $W_-$ can be obtained by solving the density matrix (Bloch equations) for Yb ion  \cite{Krysenko2023}.
For the transitions scheme in Fig.\,\ref{fig:F3}(b), where ${\bf E}_2 \perp {\bf E}_{1,3}$, we get the following expression for the total population of the excited state $^2P_{ 1/2}$ and respectively $W(\Delta)$:
\begin{equation}\label{W}
    W(\Delta) = \gamma \frac{108\, \Delta^2 \,\Omega_1^2 \,\Omega_2^2}{D} \, ,
\end{equation}
where
\begin{widetext}
\begin{eqnarray}
D &=& 2\, \Omega_1^6 +  
\left( \frac{13}{2} \Omega_2^2-48\, \delta_2 \Delta \right) \Omega_1^4 + \left(7\, \Omega_2^4 +72\,\Delta^2 \left[3\,\Omega_2^2 +\gamma^2/4 +4\,\delta_2^2 \right] \right) \Omega_1^2 \nonumber \\
&+& \frac{5}{2} \, \Omega_2^2 \left(\Omega_2^4+24\,\delta_1 \Delta\, \Omega_2^2 + \frac{288}{5}\, \Delta^2 (\gamma^2+4\,\delta_1^2)\right) +108\, \Delta^2 \,\frac{\Omega_1^2\Omega_2^2}{S_3}\, .
\end{eqnarray}
\end{widetext}
Here $S_3 = \Omega_3^2/(\gamma^2 + 4\,\delta_3^2)$ is the saturation parameter for the ${\bf E}_3$ component. Under conditions $\Omega_1 \gg \Omega_2$ and $\delta_1 \gg \gamma$ ($\delta_1 >0$), the scattering rate has the form Fig.\,\ref{fig:F4}, which is typical for EIT laser cooling technique \cite{Wineland2003}.

As can be seen from Fig.\,\ref{fig:F4}, the point $\Delta=0$ corresponds to the CPT condition. The narrow bright resonance appears to the right of CTP point corresponds to the two-photon resonance between the sublevels of the ground state. It shift from $\Delta = 0$ is determined by is ac Stark shift $\Delta_{AC}$ of dressed states Zeeman sublevels  $|^2S_{1/2}, F=1 , m=\pm 1 \rangle$ and $|^2S_{1/2}, F=0 , m=0 \rangle$. For the considered case of ${\bf E}_1$ component is chosen as a drive field, i.e. under condition $(\Omega_1, \delta_1) \gg \Omega_2$, we have 
\begin{equation}
    \Delta_{AC} = \left( \sqrt{\Omega_1^2/3+\delta_1^2} - \delta_1 \right)/2 \,.
\end{equation}
The condition 
\begin{equation}
   \Delta_{AC} = \omega_{osc} \, , 
\end{equation}
is the optimal for two-photon transitions involving to change in the vibrational number to $\Delta n = -1$, and thus results in the highest cooling rate and the lowest temperature \cite{Wineland2003,Morigi2000}. This determines the intensity of the ${\bf E}_1$ component for given $\delta_1$ and $\omega_{osc}$:
\begin{equation}
\Omega_1 = 2\,\sqrt{3} \sqrt{\omega_{osc} (\delta_1+\omega_{osc}) } \, .
\end{equation}
The obtained analytical expression (\ref{W}) allows us to estimate the limit of laser cooling of the $^{171}$Yb$^{+}$ ion in the proposed EIT scheme. In particular, the average vibrational number (\ref{Nff}) in the limit $\Omega_1 \gg \Omega_2$ takes the form
\begin{equation}
    {\bar N}_f = \frac{3\, \Omega_2^2/2+S_3 \gamma^2}{16\, \delta_1^2 \,S_3}\, ,
\end{equation}
which, for low intensity of ${\bf E}_2$ component, when  $\Omega_2 \ll \sqrt{2\,\gamma S_3/3}$, leads to
\begin{equation}\label{nbarlimit}
    {\bar N}_f = \frac{\gamma^2}{16\, \delta_1^2} \, .
\end{equation}
Therefore, deep laser cooling of $^{171}$Yb$^+$ ion down to ${\bar N}_f \ll 1$ can be achieved under condition $\delta_1  \gg \gamma$. Note that the expression (\ref{nbarlimit}) corresponds to well-known limit of EIT cooling in the standard $\Lambda$ scheme \cite{Wineland2003}.

\begin{figure}[t]
\centerline{\includegraphics[width=3.5 in]{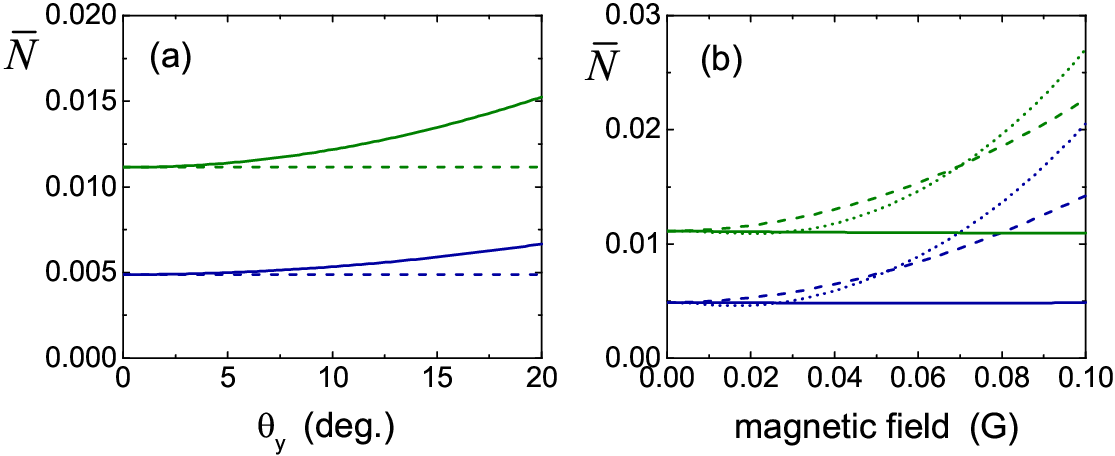}}
\caption{EIT cooling limit average quantum number  ${\bar N}$ against possible experimental imperfections: (a) as a function of angle $\theta_y$ describing deviation of ${\bf E}_2$ component light polarization  from $y$ axes (dashed lines represent results for ${\bf k}_2$ is chosen along z axis ($\theta = \pi/2$) and the solid lines represents results for $\theta = \pi/4$), and (b) as the function of residual magnetic filed (solid lines -- magnetic field along x, dashed lines -- magnetic field along y, and dotted lines -- magnetic field along z). Green (upper) lines represent results for the field parameters $\delta_1 = \delta_2 = 5 \gamma$, $\Omega_1 = 1.25 \gamma$, and blue (lower) lines for the field parameters $\delta_1 = \delta_2 = 10 \gamma$, $\Omega_1 = 1.77 \gamma$. The other  parameters $\Omega_2 = 0.1 \gamma$, $\delta_3 = 0$, $S_3 = 0.5$ are kept the same. Here $\omega_{osc}/2\pi = 600$\,kHz for considered $^{171}Yb^{+}$ trap. } \label{fig:F5}
\end{figure}

\section{Discussion}
In the following section we discuss some issues related to experimental realization of proposed scheme and it resilient against unavoidable experimental imperfections. First of all, we note that the cooling rate is determined by Lamb-Dicke parameter for two-photon transition (\ref{LD2}). It contains the Lamb-Dicke parameter for one-photon transition $\eta_1 = \sqrt{k^2 \hbar/(2M\omega_{osc})}$ and the factor depending on the angle between the wavevectors  ${\bf k}_1$ and ${\bf k}_2$. 
For our $^{171}$Yb$^+$ trap with $\omega_{osc}/2\pi \simeq 600$\,kHz, we have $\eta_1 \simeq 0.1$. Thus, in order to achieve a reasonable cooling rate, the angle $\theta$ between the wave vectors ${\bf k}_1$ and ${\bf k}_2$ (see Fig.\,\ref{fig:F2}(a)) should be large, close to $\pi/2$. 

The Fig.\,\ref{fig:F5} demonstrates EIT cooling limit ${\bar N}$ for Yb ion and it sensitivities to experimental imperfections caused by deviation of ${\bf E}_2$ component light polarization from $y$ axes and presence of residual magnetic field. As can be seen, the results are resistant to above imperfections.  Indeed, for the angle $\theta = \pi/2$ there is no dependence on deviation of ${\bf E}_2$ component light polarization from $y$ axes. For the case of $\theta = \pi/4$, the average quantum number ${\bar N}$ increases slightly, but remains enough small for a wide range of $\theta_y$.  
The residual magnetic field required for metrology clock operations is order of $0.01$\,G. As can be seen from Fig.\,\ref{fig:F5}(b), such values of the magnetic field have no effect on the EIT cooling limit. 

We emphasize, the suggested EIT cooling scheme does not require a magnetic field at all. This opens up possibilities for developing techniques for more deep control of the residual magnetic field inside the trap to the level $\sim 0.001$\,G, which potentially allows to reduce instability of optical clocks due to the second-order Zeeman shift  to the level of $10^{-19}$ and bellow. 

The cooling rate of presented EIT scheme is enough reasonable and comparable with double EIT cooling scheme implemented in \cite{Qiao}. As an example, for the field parameters corresponding to the green and blue lines on Fig.\,\ref{fig:F5} for $\theta = \pi/2$, the cooling rates are ${\dot{\bar N}} \simeq 8.2$\,ms$^{-1}$ and ${\dot{\bar N}} \simeq 4.6$\,ms$^{-1}$ respectively.

Finally, we also note that additional advantage of the presented EIT laser cooling scheme, since $\Omega_2 \ll \Omega_1$, is that the ion is localized on the lower vibrational state of the $^2S_{1/2} (F=0)$ state, which is directly used for further implementations of clock protocols with quadrupole $^2S_{1/2} (F=0) \to \,^2D_{3/2} (F=2)$ or octupole $^2S_{1/2} (F=0) \to\, ^2F_{7/2} (F=3)$ transitions.

\section{Conclusion}
We propose EIT scheme for ground state laser cooling of $^{171}$Yb$^{+}$ ions that does not require the use of a magnetic field. For laser cooling, a polychromatic configuration of the light field is used, consisting of three monochromatic running waves resonant to optical transitions of the $^2S_{1/2} \to\, ^2P_{1/2}$ line. For the first stage of Doppler cooling, the light frequency components are running waves with codirectional linear polarizations. 
In this case, each of the frequency components of the field has a mechanical action on the ion, which finally leads to the cooling down to temperature of the Doppler limit. For the trap with typical oscillation frequency is about $600$ kHz, this temperature corresponds to the average vibrational quantum number ${\bar N} \simeq 20$. To implement the second stage of deep laser cooling down to the motional ground state, i.e. ${\bar N} \ll 1$, the polarization of one of the frequency components have to be oriented relative to the others by an angle of $90^\circ$. 

On the one hand, the exclusion of the magnetic field from laser cooling allows to reduce the total time of ``clock operation'' cycle by eliminating the time interval required to turn off and attenuate the magnetic field, which is used in standard two-frequency field cooling scheme. Indeed, reducing the cycle time contributes to faster accumulation of measurement statistics in optical frequency standards. On the other hand, the absence of the need to use a magnetic field for cooling process allows for more accurate control of the residual magnetic field and minimize its fluctuations in various measurement cycles, which is important for further improving the accuracy and long-term stability of optical atomic clocks based on $^{171}$Yb$^{+}$. In addition, deep ground state cooling to ${\bar N} < 1$ significantly suppress the second-order Doppler shift to a level bellow $\Delta \nu/\nu < 10^{-19}$, which allows to remove it from the uncertainty budget of frequency standards based on $^{171}$Yb$^{+}$.

Note, that suggested method of the ground-state cooling is also important  for quantum logic elements based on cold $^{171}$Yb$^{+}$ ions.

We thank Nils Huntemann for useful discussions.
The research was supported by the grant from the Russian
Science Foundation (project no. 23-22-00198), https://rscf.ru/project/23-22-00198/


\nocite{*}


\begin{references}
\bibitem{Cornell2002} E. Cornell and C.E. Wieman Rev. Mod. Phys. {\bf 74,} 875 (2002)
\bibitem{Ketterle} W. Ketterle  Rev. Mod. Phys. {\bf 74,} 1131 (2002)
\bibitem{DeMarco} B. DeMarco and D.S. Jin Science {\bf 285,} 1703–1706 1999
\bibitem{DeMarco_PRL} B. DeMarco B, J.L. Bohn, J.P. Burke, M. Holland,  and D.S. Jin, Phys. Rev. Lett. {\bf 82,} 4208–4211 (1999)

\bibitem{Nielsen} M. A. Nielsen, I.L. Chuang ``Quantum Computation and Quantum Information'', Cambridge University Press, 2010

\bibitem{Falke} S. Falke, et al. New J. Phys. {\bf 16,} 073023 (2014)
\bibitem{Katori2020} M. Takamoto, I. Ushijima, N. Ohmae, T. Yahagi, K. Kokado, H. Shinkai, and H. Katori, Nat. Photonics {\bf 14,} 411–415 (2020)
\bibitem{McGrew} W. F. McGrew, X. Zhang, R. J. Fasano, S. A. Schaffer, K. Beloy, D. Nicolodi, R. C. Brown, N. Hinkley, G. Milani, M. Schioppo, T. H. Yoon, A. D. Ludlow, Nature {\bf 564,} 87–90 (2018) 

\bibitem{Wineland_Al} C. W. Chou, D. B. Hume, J. C. J. Koelemeij, D. J. Wineland, T. Rosenband, Phys. Rev. Lett. {\bf 104,} 070802 (2010)
\bibitem{Huntemann} N. Huntemann, C. Sanner, B. Lipphardt, C. Tamm, and E. Peik, Phys. Rev. Lett. {\bf 116,} 063001 (2016)
\bibitem{Huang_Ca} Y. Huang, H. Guan, P. Liu, W. Bian, L. Ma, K. Liang, T. Li, and K. Gao, Phys. Rev. Lett. {\bf 116,} 01300 (2016)

\bibitem{Lion2017} G. Lion, I. Panet, P. Wolf, C. Guerlin, S. Bize, and P. Delva,  Journal of Geodesy {\bf 91,} 597–611 (2017)
\bibitem{Ludlow2018} W. F. McGrew, X. Zhang X, R. J. Fasano, S. A. Schaffer, K. Beloy, D. Nicolodi, R. C. Brown, N. Hinkley, G. Milani, M. Schioppo, T.H. Yoon, and A.D. Ludlow, Nature {\bf 564,} 87–90, (2018) 

\bibitem{Godun14} R.M. Godun. P.B.R. Nisbet-Jones, J.M. Jones, S.A. King, L.A.M. Johnson, H.S. Margolis, K. Szymaniec, S.N. Lea, K. Bongs, and P. Gill, Phys. Rev. Lett. {\bf 113,} 210801 (2014)
\bibitem{Huntemann14} N. Huntemann, B. Lipphardt, Chr. Tamm, V. Gerginov, S. Weyers, and E. Peik, Phys. Rev. Lett. {\bf 113,} 210802 (2014)


\bibitem{Dzuba} V. Dzuba, V.V. Flambaum, M.S. Safronova, S.G. Porsev, T. Pruttivarasin, M.A. Hohensee, and H. Haffner, Nature Physics {\bf 12,} 465–468 (2016)
\bibitem{Sanner} C. Sanner, N. Huntemann, R. Lange, C. Tamm, E. Peik, M.S. Safronova and S. G. Porsev,  Nature {\bf 567,} 204–208 (2019)
\bibitem{Laura} L.S. Dreissen, C.-H. Yeh, H. A. Fürst, K.C. Grensemann,  T.E. Mehlst\"{a}ubler, Nature Communications {\bf 13,} 7314 (2022)

\bibitem{Arvanitaki} A. Arvanitaki, J. Huang, and K.V. Tilburg, Phys. Rev. D {\bf 91,} 015015 (2015)
\bibitem{Stadnik} Y.V. Stadnik,  V.V. Flambaum, Phys. Rev. Lett. {\bf 115,} 201301 (2015)


\bibitem{Tamm} Chr. Tamm, S. Weyers, B. Lipphardt, and E. Peik, Phys. Rev. A {\bf 80,} 043403 (2009)
\bibitem{Prudnikov_2017} O.N. Prudnikov, S.V. Chepurov, A.A. Lugovoy, K. M. Rumynin, S.N. Kuznetsov, A. V. Taichenachev, V.I. Yudin, S.N. Bagayev, Quantum Electronics, {\bf 47,} 806–811 (2017)
\bibitem{Prudnikov_2019} S.V. Chepurov, A.A. Lugovoy, O.N. Prudnikov, A.V. Taichenachev, S.N. Bagayev, Quantum Electronics {\bf 49,}  412 – 417 (2019) 

\bibitem{Kolachevsky} M. A. Aksenov, I. V. Zalivako, I. A. Semerikov, A. S. Borisenko, N. V. Semenin, P. L. Sidorov, A. K. Fedorov, K. Yu. Khabarova, and N. N. Kolachevsky
Phys. Rev. A {\bf 107,} 052612 (2023)

\bibitem{Wineland} D.J.  Wineland, W.M. Itano, Phys. Rev. A  {\bf 20,} 1521 (1979)  
\bibitem{Javanainen1981} J. Javanainen, Appl. Phys. {\bf 24,} 151-162 (1981)
\bibitem{Wineland2003} D. Leibfried, R. Blatt, C. Monroe, D. Wineland, Reviews Of Modern Physics {\bf 75,} 281 (2003)

\bibitem{Morigi2000} G. Morigi, J. Eschner, and C. H. Keitel, Phys. Rev. Lett. {\bf 85,} 4458 (2000).
\bibitem{Morigi2003} J. Eschner, G. Morigi, F. Schmidt-Kaler, and R. Blatt, J. Opt. Soc. Am. B {\bf 20,} 1003 (2003)
\bibitem{Roos2016} R. Lechner, C. Maier, C. Hempel, P. Jurcevic, B.P. Lanyon, T. Monz, M. Brownnutt, R. Blatt, and C.F. Roos, Phys. Rev. A  {\bf 93,} 053401 (2016)
\bibitem{Khabarova} I. A. Semerikov, I. V. Zalivako, A. S. Borisenko, K. Y. Khabarova, and N. N. Kolachevsky, Journal of Russian Laser Research {\bf 39,} 568, (2018)

\bibitem{Evers} J. Evers and C. H. Keitel, Europhys. Lett., 68, 370 (2004).

\bibitem{Qiao} Mu Qiao, Ye Wang, Zhengyang Cai, Botao Du, Pengfei Wang, Chunyang Luan, Wentao Chen, Heung-Ryoul Noh, and Kihwan Kim, Phys. Rev.  Lett. {\bf 126,} 023604 (2021).

\bibitem{Krysenko2023} D.S. Krysenko, O.N. Prudnikov, JETP {\bf 137,} 239-245 (2023)

\end{references}

\end{document}